# ERIOS: Co-construction of a Dynamic Temporal Visualization Tool in the Electronic Health Record

# ERIOS : Co-construction d'un outil de visualisation temporelle dynamique dans le Dossier Patient Informatisé

Robert, Louise

ERIOS, **E**space de Recherche et d'**I**ntégration des **O**utils **N**umériques en **S**anté, CHU de Montpellier, louise.robert@chu-montpellier.fr

Luzurier, Quentin

ERIOS, **E**space de Recherche et d'**I**ntégration des **O**utils **N**umériques en **S**anté, Dedalus, quentin.luzurier@dedalus.eu

Velcker, Anaïs

ERIOS, **E**space de Recherche et d'**I**ntégration des **O**utils **N**umériques en **S**anté, Dedalus, Polytech Montpellier, Université de Montpellier, anais.velcker@dedalus.eu

Mathieu, Emma

ERIOS, **E**space de Recherche et d'**I**ntégration des **O**utils **N**umériques en **S**anté, Polytech Montpellier, Université de Montpellier, emma.mathieu@etu.umontpellier.fr

Fontaine, Loïc

ERIOS, **E**space de Recherche et d'**I**ntégration des **O**utils **N**umériques en **S**anté, Dedalus, loic.fontaine@dedalus.eu.

Morquin, David

ERIOS, **E**space de Recherche et d'**I**ntégration des **O**utils **N**umériques en **S**anté, CHU de Montpellier, d morquin@chu-montpellier.fr

ERIOS, is a collaborative project between Dedalus, a health software company, Montpellier University Hospital Center (CHU), and the University of Montpellier. This initiative aims to incorporate research and development (R&D) directly within the hospital, focusing on co-creating components of the Electronic Health Record (EHR) alongside end-users. The project was initiated with two initial use cases, which led to the development of components for dynamic temporal visualization, now integrated into specific dashboards. The application of academic recommendations regarding user engagement methodology and human-computer interactions significantly enhanced our ability to meet user needs.

ERIOS, Espace de Recherche et d'Intégration des Outils Numériques en Santé, est une collaboration entre Dedalus, éditeur de logiciels de santé, le CHU de Montpellier, et l'Université de Montpellier. Cette initiative vise à intégrer la recherche et le développement (R&D) au sein même du CHU de Montpellier, en co-construisant des composants du Dossier Patient Informatisé (DPI) avec les utilisateurs finaux. Le projet a débuté avec deux premiers cas d'usage, qui ont permis le développement de composants pour la visualisation temporelle dynamique, intégrés dans des tableaux de bord spécifiques. L'application des recommandations académiques en termes de méthodologie d'implication des utilisateurs finaux et d'interactions homme-machine a significativement amélioré la réponse aux besoins des utilisateurs.

**Mots-clés additionnels :** Dossier Patient Informatisé, Interactions Homme-machine, Visualisation temporelles dynamiques, Tableaux de bord.

**ACM Reference Format :**

## 1 INTRODUCTION

ERIOS, Espace de Recherche et d'Intégration des Outils Numériques en Santé est une initiative conjointe de l'éditeur de logiciels de santé Dedalus, du CHU de Montpellier, et de l'Université de Montpellier. Le consortium a installé une équipe de R&D au cœur du CHU de Montpellier avec pour objectif de co-construire avec les utilisateurs finaux des composants du Dossier Patient Informatisé (DPI). Nous avons mené deux cas premiers cas d'usage en 2023 : « IsoPsy » dédié au suivi de l'isolement thérapeutique en psychiatrie et « AtbViz » pour la gestion des séquences de traitements anti-infectieux. En s'inscrivant dans une approche de design participatif [Sanders, Stappers 2008] nos développements sont fondés sur une analyse rigoureuse des besoins exprimés par les utilisateurs, couplée à une étude approfondie des travaux académiques sur la visualisation et la structuration des données au sein des DPI. Cela nous a conduit à développer des composants de visualisations temporelles dynamiques. Ils sont intégrés dans des tableaux de bord (TDB) spécifiques, conçus pour les tâches déterminées de nos cas d'usage.

## 2 LES DOSSIERS PATIENTS INFORMATISES

Introduits au début des années 2000, les DPI sont aujourd'hui incontournables. Ils sont le pilier de toutes les activités hospitalières et du parcours de soins des patients. Bien que l'apport de ces systèmes soit indéniable, leur conception présente encore de nombreux défis. En particulier, il a été souligné que de nouvelles modalités de visualisation ont le potentiel d'améliorer la charge cognitive des professionnels de santé lors des prises de décisions médicales et permettre une meilleure coordination des équipes. Un autre enjeu est qu'une charge trop importante de documentation dans le logiciel peut conduire à sentiment de déqualification de certains utilisateurs et se faire au détriment de la dimension relationnelle des soins. Il a été démontré qu'un médecin hospitalier passe en moyenne 5,3 heures par jour à utiliser son DPI, contre seulement 1,3 heure au contact direct des patients [Morquin, Ologeanu-Taddei 2016], soit environ 37 % de son temps quotidien de travail effectif devant un ordinateur [Shanafelt et al. 2016]. De plus, l'attention portée à l'écran de l'ordinateur pendant la consultation peut créer une barrière perçue entre le médecin et le patient, altérant la qualité de l'interaction. Cette situation entraîne parfois un sentiment de déshumanisation des soins pour les patients, qui peuvent se sentir moins écoutés ou compris. Ces éléments peuvent affecter la perception de la qualité des soins et diminuer la confiance du patient envers son soignant.

## 3 NOS CAS D'USAGE : IDENTIFICATION DE LA SITUATION ACTUELLE ET DES BESOINS DES UTILISATEURS

Notre premier cas d'usage (IsoPsy) concerne une procédure médico-administrative en application d'un article de loi[1] pour le suivi légal des patients en isolement thérapeutique en psychiatrie. Les utilisateurs nous avaient fait part de leurs difficultés à coordonner leurs tâches dans un processus de travail complexe et fortement contraint par le temps. C'est une véritable course contre la montre, impliquant de nombreux acteurs et ayant des conséquences significatives si les tâches ne sont pas achevées dans un délai strict. Actuellement, le processus de travail s'avère inefficace, principalement en raison d'un suivi des tâches qui se fait de manière fragmentée et dispersé à travers de multiples canaux de communication tels que les transmissions orales, les courriels, les documents papier, les tableaux Excel, les affichages muraux dans les services et les appels téléphoniques. Les utilisateurs ont exprimé le besoin de comprendre, à tout moment, comment organiser temporellement les tâches collectives et individuelles. Ils souhaitaient une fonction de priorisation ainsi que la possibilité de comprendre quelles tâches doivent être anticipées en fonction du calendrier réel de l'organisation des soins.

---

[1] Article 17 de la LOI n° 2022-46 du 22 janvier 2022.

Notre second cas d'usage (AtbViz) concerne la gestion des traitements anti-infectieux. Dans le processus de prescription thérapeutique, les médecins sont confrontés à la nécessité de corréler une multitude de données variées, en tenant compte de leur évolution temporelle. Actuellement, l'intégration des données dépend largement de l'utilisateur, qui doit parcourir divers onglets pour appréhender la situation complète du patient et anticiper l'évolution future. Pour évaluer la situation d'un patient et décider de l'approche thérapeutique en matière d'anti-infectieux, un infectiologue doit faire de nombreux allers-retours dans le DPI, il faut consulter en moyenne 70 écrans pour un seul cas. Les utilisateurs ont exprimé le besoin de corréler plus facilement des données dans le temps avec un agencement visuel de ces données qui soit adapté à leurs modalités de prise de décision, sous formes de composants distincts d'un même tableau de bord.

**4 OPERATIONNALISATION DES THEORIES IHM DANS LE DPI**

Nous avons tout d'abord fait l'hypothèse que consulter les données nécessaires sur un seul écran éviterait aux utilisateurs de devoir passer au crible de nombreux onglets. Il a en effet été souligné que l'agencement actuel des données dans le DPI n'est pas pertinente au regard des associations cliniques, les données associées étant présentées dans des écrans distincts [Senathirajah et al. 2017]. Le concept de « problème de fragmentation de l'affichage » nous permettait alors de mieux comprendre la fatigue informationnelle des utilisateurs qui doivent consulter de nombreux écrans et retenir des informations plutôt que de pouvoir visualiser toutes les informations pertinentes ensemble.

Nous nous sommes également appuyés sur les travaux portant sur l'adéquation tâche-donnée (*task-data*) dans les systèmes de visualisation. La théorie de l'ajustement cognitif [*cognitive fit,* Vessey 1991] suggère que la performance dans la résolution de problèmes dépend de l'interaction entre la représentation de l'information et la tâche de résolution de problème elle-même. Le niveau de congruence entre la tâche à accomplir et la manière dont les informations sont présentées influence positivement la performance de l'humain dans la résolution de problèmes. Ainsi, la littérature nous permettait de faire une autre hypothèse : une forte densité de données pourrait aider à résoudre les problèmes de navigations répétitives et d'assimilation complexe d'informations éparpillées dans le dossier des patients [Ghalayini et al. 2020]. Par la pratique du design participatif nous avons cherché à éviter tout « encombrement » des données [*Clutter* ; Moacdieh et al. 2015]. Il y'a en effet consensus de longue date sur l'importance de la participation des utilisateurs au déploiement des systèmes d'information en santé [Grindell et al. 2022]. Considérant que les problèmes d'utilisabilité des DPI ne sont pas attribuables à une trop grande quantité d'informations mais à leur agencement, nous avons organisé des ateliers afin que les utilisateurs finaux organisent eux-mêmes les données pertinentes et nous indiquent de quelle manière ils voudraient les visualiser.

Enfin, un autre apport est celui de la théorie de la conscience situationnelle [*situationnal awareness* ; Endsley 1995] qui est un cadre conceptuel intéressant pour penser la prise de décision médicale dans le DPI. Comme recommandé par Chang.et al. (2017) nous nous sommes assurés que dans nos composants du DPI, la réflexion peut être est décomposée en trois niveaux. Le premier niveau est celui de la perception (détection et identification des informations pertinentes). Au deuxième niveau, l'utilisateur va intégrer et interpréter les informations perçues. Enfin, le troisième niveau est celui de la projection : c'est le niveau final où un individu est capable d'anticiper les futurs états de la situation fondés sur la compréhension actuelle de la situation.

**5 PARAMETRAGE ET FONCTIONALITES DES VISUALISATIONS CHRONOLOGIQUES ADAPTEES AUX SPECIFICITES DES CAS D'USAGE ET DES ACTEURS**

Le cadre de développement ERIOS se focalise sur l'implémentation de nouveaux composants ou de nouvelles fonctionnalités de visualisation avec la solution Dedalus Care4u. Cette suite logicielle full web comprend une application de visualisation de données du patient, du service ou de l'utilisateur sous forme de TDB. Nommée Composer U, cette application permet de construire des TDB de manière modulable avec différents composants préconfigurés. Notre équipe de développement avait pour objectif d'adapter ou de développer de nouveaux composants répondant aux besoins métiers identifiés mais pour lesquels aucune solution suffisamment adaptée n'existe encore. S'inscrivant dans cette démarche, les deux cas d'usage ERIOS menés en 2023 nécessitaient la conception d'un composant de visualisation temporelle, chronologique, dynamique et interactive d'une quantité importante d'informations.

Après analyse des besoins des utilisateurs finaux, des processus métiers et de la littérature scientifique, l'utilisation d'un composant de type timeline répondait le mieux aux spécifications métiers. Ce nouveau composant a été prototypé en utilisant une bibliothèque existante (vis.js) avec des adaptions spécifiques aux besoins des cas d'usage ERIOS. Ces adaptations ont été notamment réalisées par ajout de fragments de code JavaScript directement dans l'interface d'intégration du Composer U, cet outil ayant été conçu pour recevoir de telles surcharges.

Ce processus d'intégration a rapidement permis de mettre en place des fonctionnalités générales du composant timeline, dans le respect des règles d'expérience utilisateur Care4u. Il s'agissait notamment des fonctionnalités de navigation avec possibilité de défilement non limité dans le temps et une fonctionnalité de zoom et dézoom temporel. La navigation devant être simple, affordante et accessible, celle-ci a été permise soit par clic sur des boutons de navigation situés sur la timeline, soit directement par manipulation de type « drag » (déplacement avec maintien du clic) pour le défilement et « scroll » (molette de la souris) pour le zoom / dézoom.

## 6   ISOPSY : UNE VISUALISATION TEMPORELLE AVEC FONCTION DE PRIORISATION ET FILTRE POUR ANTICIPATION DES TACHES

L'enjeu majeur était de situer chronologiquement les différentes tâches, de pouvoir facilement repérer une échéance proche et son éventuelle nécessité d'anticipation (en cas de week-end ou de jour férié notamment). L'identification du professionnel à qui incombait chaque tâche était également une demande forte des utilisateurs.

Dans le prototype conçu, chaque tâche est représentée par un item rectangulaire sur le composant timeline (fig.1) selon un code couleur allant du rouge au vert en fonction du délai restant pour accomplir la tâche. Des informations plus détaillées sont facilement visualisables par un survol de l'item via une infobulle. L'outil ayant pour objectif de permettre une visualisation partagée par tous les professionnels prenant part au processus d'isolement thérapeutique, le choix a été fait de concevoir un seul et même TDB pour tous les métiers, avec un système optionnel de filtre par métier. Un menu déroulant permet ainsi de filtrer les tâches pour n'afficher que celles correspondant au métier sélectionné.

La solution proposée pour l'anticipation des tâches a tout d'abord été de représenter les week-end et les jours fériés sur fond gris sur la timeline. Dans un second temps, une option accessible par un clic sur un bouton de la timeline a été ajoutée pour permettre de recalculer les échéances en tenant compte des règles d'anticipation précisées. Afin de faciliter la validation des tâches, des interactions possibles directement depuis la timeline ont été implémentées. Un double-clic sur l'item de la tâche sur la timeline suffit à ouvrir la vue de validation, avec mise à jour automatique de l'ensemble du tableau de bord. Ces fonctionnalités ont été mises en place dans des TDB à l'échelle du dossier d'un patient (fig.1), mais également du service (fig.2) ou de l'ensemble de l'établissement. Pour ces derniers TDB, l'identité du patient et l'indication du service dans lequel il était hospitalisé complètent les informations utiles à la gestion des isolements thérapeutiques.

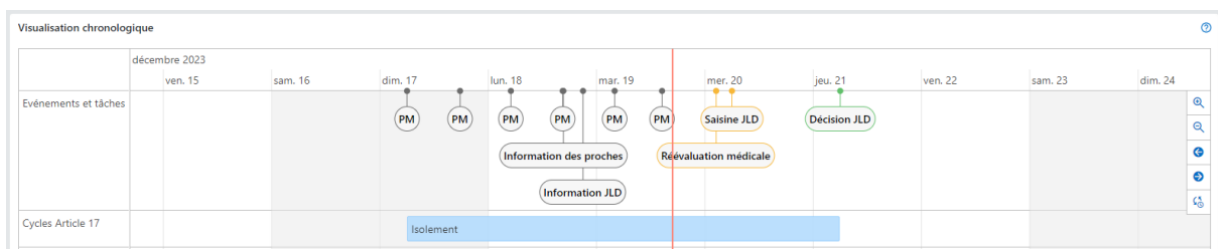

Figure 1. Composant timeline inséré dans le tableau de bord du dossier patient (cas d'usage « IsoPsy »)

PM : Prescription médicale ; JLD : Juge des Libertés et de la Détention

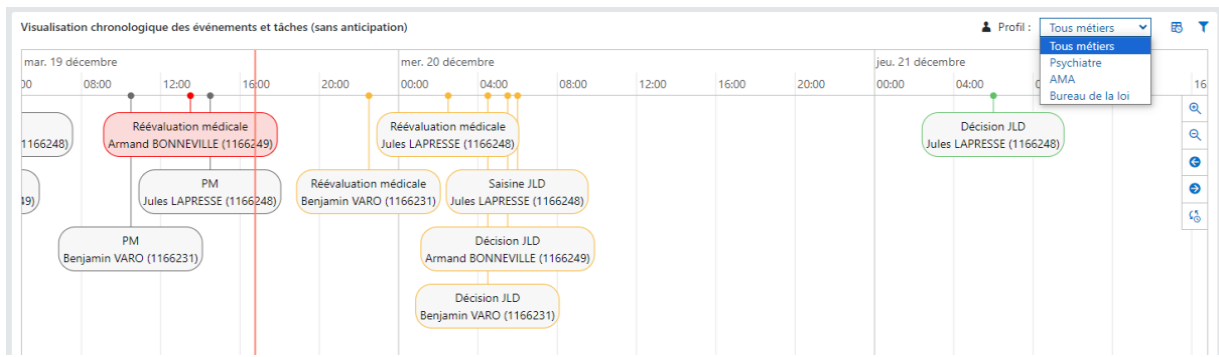

Figure 2. Composant timeline inséré dans le tableau de bord du service, avec filtre optionnel par métier (cas d'usage « IsoPsy »)

PM : Prescription médicale ; JLD : Juge des Libertés et de la Détention

## 7 ATBVIZ : DES VISUALISATIONS TEMPORELLES DYNAMIQUES ET SYNCHRONISEES AVEC UNE SELECTION DES DONNEES PERTINENTES, DE LEUR MODE DE VISUALISATION, DES VALEURS, EXPRIMEES PAR LES UTILISATEURS

Pour notre deuxième cas d'usage (AtbViz), l'enjeu était de visualiser une quantité importante d'informations de nature différente (données de prescription, de biologie, constantes cliniques, annotations) selon un regroupement thématique pertinent. Cet agencement devait permettre aux utilisateurs d'accomplir leurs différentes tâches : suivi des thérapeutiques, de l'efficacité, des investigations microbiologiques, de la tolérance des traitements. La solution a alors été de présenter sur un même TDB plusieurs composants sous forme de timelines ou de graphiques (un composant pour chacun des regroupements thématiques), avec une synchronisation temporelle : la manipulation de l'un des composants s'applique immédiatement sur tous les autres. La nécessité de pouvoir naviguer depuis les différents composants (timelines et graphiques), exprimé dès les premiers retours utilisateurs, a également été implémentée avec utilisation soit des boutons de navigation, soit directement par drag ou scroll.

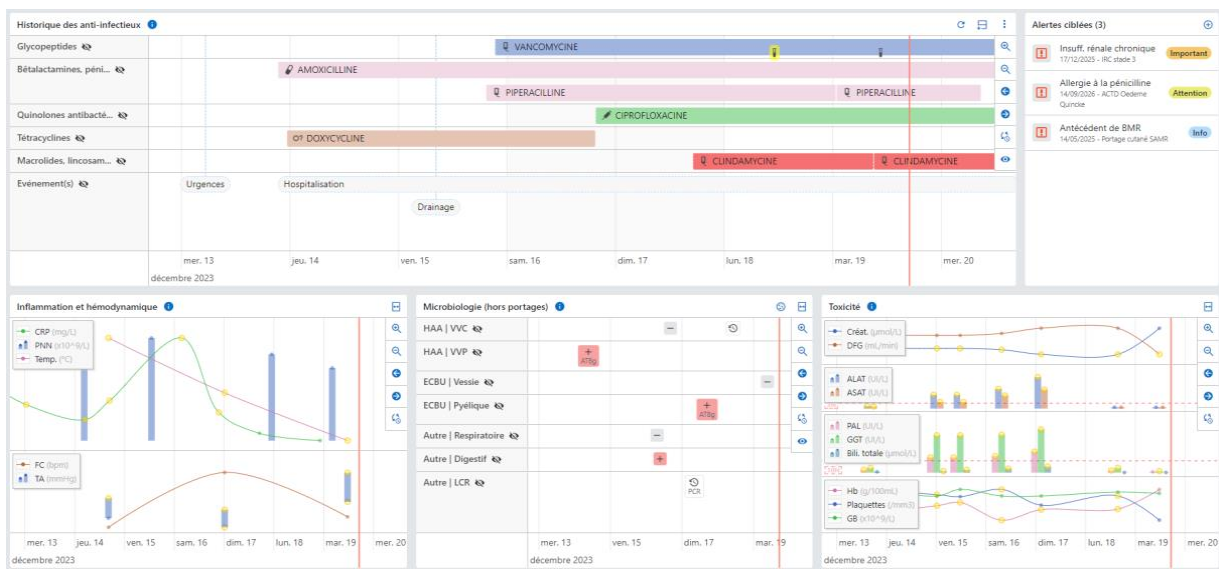

Figure 3. Composants timeline et graphiques synchronisés insérés dans le tableau de bord du patient (cas d'usage « AtbViz »)

## 8 LIMITES

L'expérimentation ERIOS confirme que la collaboration étroite entre les utilisateurs finaux et les développeurs au sein de l'établissement de santé optimise la conception et l'amélioration des composants du DPI. Les intégrations réalisées pour ces deux premiers cas d'usage ont cependant mis en évidence certaines difficultés. Une des principales consistait en la diversité des demandes des utilisateurs finaux, parfois contradictoires. Le choix des solutions à intégrer devait également tenir compte des contraintes techniques, des directives vis-à-vis de l'expérience utilisateur, ainsi que des considérations de qualité produit et de sécurité. Nous avons mis en place des moyens de sélection des solutions les plus efficientes à l'aide d'outil d'analyse des ateliers utilisateurs et de réunions d'arbitrage pour statuer collégialement sur ces choix.

Du point de vue plus technique, les limites liées à l'interopérabilité entre les systèmes d'information ont complexifié les développements et les tests utilisateurs. L'outil utilisé pour développer les nouveaux composants de visualisation (DPI Dedalus Care4u) n'était pas intégré nativement au système d'information actuellement en place au CHU de Montpellier (DPI Dedalus DxCare). Les prototypes de visualisation conçus nécessitant des données médicales cohérentes actualisées en continu, les tests utilisateurs en conditions réelles s'en sont trouvés complexifiés. Une alimentation de la base de données de l'environnement de développement et de test avec des données de production de l'établissement hospitalier s'est révélé être un challenge majeur. Afin de ne pas limiter les tests utilisateurs, des solutions de contournement ont été mises en place pour simuler ces données, dans l'attente de leur disponibilité dans le produit final. Le Composer U permet en effet d'ajouter des données, principalement sous format JSON, permettant dans un premier temps l'utilisation d'informations factices. Pour résoudre ces problématiques d'interopérabilité, des API RESTful respectant les normes de données de santé sont en cours de développement et faciliteront prochainement l'intégration des nouveaux composants dans le DPI actuellement en place.

Dans le cadre du prototypage technique des nouveaux composants, d'autres contraintes liées à l'utilisation d'une bibliothèque tierce (vis.js) ont été rencontrées. L'intégration dans Care4u limitait l'utilisation de certaines fonctionnalités natives de cette bibliothèque et des contournements techniques ont été réalisés pour répondre aux spécifications définies avec les utilisateurs finaux. Néanmoins, les développements ont été réalisés pour garantir au mieux l'évolutivité et l'applicabilité de ces nouveaux composants à d'autres établissements de santé. Dans la suite de cette étape de prototypage menée par ERIOS, le processus de développement se poursuivra pour intégration au produit final avec une prise en charge et une correction de ces limitations techniques.

**9   PERSPECTIVES**

Dans la phase actuelle du projet, nous menons des tests avec les utilisateurs finaux au CHU de Montpellier en adaptant le cadre d'évaluation de tableaux de bord proposé par Zhuang et al. [2022] aux tâches attendues de nos composants. Par la suite, nous explorerons les visualisations temporelles dynamiques dans d'autres cas d'usages. Il y'a en effet une importance universelle de la compréhension des processus temporels en médecine, influençant de manière significative la prise de décision médicale, l'organisation des soins et la coordination des parcours des patients. Nous envisageons également d'approfondir d'autres besoins exprimés par les utilisateurs dans ces deux premiers cas d'usage, d'une part les fonctionnalités d'annotation, de l'autre la visualisation de données manquantes dans les visualisations temporelles dynamiques.

**10 CONCLUSION**

En nous appuyant sur la littérature afin d'améliorer les modalités de consultation des données, pour mieux répondre aux besoins réels des utilisateurs et aux dynamiques des processus de travail, notre objectif est de développer un DPI qui accroîtra significativement la satisfaction des professionnels de santé dans leur pratique quotidienne. Cet outil devrait contribuer à la réduction du stress, libérer du temps précieux pour des interactions de qualité avec les patients. L'objectif étant sur le long terme de veiller à préserver la qualité de notre système de santé, des parcours des patients et des conditions de travail des personnels hospitaliers dans des environnements de plus en plus digitalisés.

# Références

Al Ghalayini, M., Antoun, J., & Moacdieh, N. M. (2020). Too much or too little? Investigating the usability of high and low data displays of the same electronic medical record. Health informatics journal, 26(1), 88–103. https://doi.org/10.1177/1460458218813725

Chang, Bora et al. "Triangulating Methodologies from Software, Medicine and Human Factors Industries to Measure Usability and Clinical Efficacy of Medication Data Visualization in an Electronic Health Record System." AMIA Joint Summits on Translational Science proceedings. AMIA Joint Summits on Translational Science vol. 2017 473-482. 26 Jul. 2017

Endsley, M. R. (1995). Toward a theory of situation awareness in dynamic systems. Human Factors, 37(1), 32–64. https://doi.org/10.1518/001872095779049543

Grindell, C., Coates, E., Croot, L., & O'Cathain, A. (2022). The use of co-production, co-design and co-creation to mobilise knowledge in the management of health conditions: a systematic review. BMC health services research, 22(1), 877. https://doi.org/10.1186/s12913-022-08079-y

Moacdieh N, Sarter N. (2015). Clutter in electronic medical records: examining its performance and attentional costs using eye tracking. Hum Factors, 57(4), 591–606.

Morquin, D., & Ologeanu-Taddei. (2016). Professional Facing Coercive Work Formalization: Vicious Circle of the Electronic Medical Record (EMR) Implementation and Appropriation. Procedia Computer Science, 100, 652-657. https://doi.org/10.1016/j.procs.2016.09.207

Sanders, E.B.N. & Stappers, P.J. (2008). Co-creation and the new landscapes of design. CoDesign: International Journal of CoCreation in Design and the Arts, 4(1), 5-18.

Senathirajah, Y., Wang, J., Borycki, E., & Kushniruk, A. (2017). Mapping the Electronic Health Record: A Method to Study Display Fragmentation. Studies in health technology and informatics, 245, 1138–1142.
Senathirajah and S. Bakken, (2011). Visual clustering analysis of CIS logs to inform creation of a user-configurable Web CIS interface. Methods Inf Med, 50, 337-348.

Shanafelt, T. D., Dyrbye, L. N., Sinsky, C., Hasan, O., Satele, D., Sloan, J., & West, C. P. (2016). Relationship Between Clerical Burden and Characteristics of the Electronic Environment With Physician Burnout and Professional Satisfaction. Mayo Clinic Proceedings, 91(7), 836-848. https://doi.org/10.1016/j.mayocp.2016.05.007

Vessey, I. (1991). Cognitive fit: A theory-based analysis of the graphs versus tables literature. Decision Sciences, 22(2), 219–240.

Zhuang, M., Concannon, D., & Manley, E. (2022). A framework for evaluating dashboards in healthcare. IEEE Transactions on Visualization and Computer Graphics, 28(4), 1715-1731. https://ieeexplore.ieee.org/document/9721816/